\documentclass[conference]{IEEEtran}

\ifCLASSINFOpdf
\else
\fi
\usepackage{graphicx}
\usepackage{amsmath,amssymb}
\usepackage{tikz}
\usetikzlibrary{arrows.meta,fit,positioning,shapes.geometric,shapes.symbols}
\usepackage{xcolor}
\usepackage{booktabs}
\usepackage{multirow}
\usepackage{array}
\newcolumntype{L}[1]{>{\raggedright\arraybackslash}p{#1}}
\usepackage{algorithm}
\usepackage{algpseudocode}
\usepackage{listings}
\lstdefinelanguage{yaml}{keywords={provider,version,references,steps,module,weight,rules,check,score_impact,threshold,reference_key},
  keywordstyle=\color{tpblue}\bfseries, basicstyle=\ttfamily\scriptsize, comment=[l]{\#},
  commentstyle=\color{gray}, showstringspaces=false}
\definecolor{tpblue}{RGB}{26,83,130}
\definecolor{tpcyan}{RGB}{42,151,178}
\definecolor{tpgreen}{RGB}{45,126,96}
\definecolor{tporange}{RGB}{210,119,35}
\definecolor{tplight}{RGB}{241,246,249}
\tikzset{
  tpbox/.style={draw=tpblue, rounded corners=2pt, fill=tplight,
    align=center, minimum height=6mm, inner sep=3pt, font=\scriptsize},
  tpdata/.style={draw=tpgreen, rounded corners=2pt, fill=tpgreen!8,
    align=center, minimum height=6mm, inner sep=3pt, font=\scriptsize},
  tparrow/.style={-{Latex[length=2mm]}, semithick, draw=tpblue},
  tpdash/.style={-{Latex[length=2mm]}, semithick, dashed, draw=tporange}
}
\tikzset{person/.pic={
    \fill[#1] (0,0.42) circle (0.13);
    \fill[#1] (-0.22,0) arc (180:0:0.22 and 0.28) -- cycle;}}
\newcommand{\cmark}{$\checkmark$}
\newcommand{\xmark}{--}
\begin{document}

\title{TrustPaper for forged PDF with Issuer Fingerprints}

\author{\IEEEauthorblockN{Yassine EL JAKANI, Mohamed Nadir RHAZI, Othmane MEKNASSI, Oualid CHBIKI,\\ Mounira MSAHLI, François GONNET}
\IEEEauthorblockA{Télécom Paris, Institut Polytechnique de Paris, UBG.Cyber}
\IEEEauthorblockA{(yassine.eljakani, mohamed.rhazi, othmane.meknassi, oualid.chbiki, mounira.msahli)@telecom-paris.fr\\ francois@ubg-interactive.fr}\\}

\maketitle

\begin{abstract}
Sick leave certificates are increasingly delivered as PDF files by online teleconsultation platforms, and can be altered in minutes with free editors. Generic PDF scanners report how a file was built, but cannot tell whether that construction is legitimate: an incremental update, an embedded font or a given producer string is normal for one issuer and anomalous for another. We present TrustPaper solution, an issuer-aware forensic approach that combines three elements: the claimed issuer of the document, a reference fingerprint learned beforehand from genuine documents of that issuer, and an explainable combination of evidence from several forensic modules. Fingerprints are stored as versioned declarative profiles, and every anomaly keeps its observed value, expected value and triggering rule. We formalize the threat model, the fingerprint and the scoring model, and evaluate the open-source prototype on a controlled corpus documents from two issuers and six forgery techniques. With the issuer fingerprint, all forgeries made by opportunistic and metadata-aware attackers are flagged with no false positive on genuine test documents. 
\end{abstract}

\begin{IEEEkeywords}
PDF forensics, document fraud, explainable detection.
\end{IEEEkeywords}

\IEEEpeerreviewmaketitle

\section{Introduction}

In France, online teleconsultation platforms, operated as Software-as-a-Service (SaaS), deliver sick leave certificates as PDF files, and official forms can be filled on a computer. Meanwhile, the public health insurance reports a strong increase of sick-leave fraud, largely based on edited or recreated documents \cite{1}.

A PDF can be modified through incremental updates, metadata rewriting or embedded fonts \cite{2}, operations that free online editors make available to anyone. On the defensive side, existing tools mostly verify a digital signature when one exists, and previous work showed that even signed PDFs can be manipulated without breaking the signature \cite{3}, \cite{4}. Although the literature contains solutions based on signatures or blockchain \cite{8}, \cite{9} most certificates are not signed, and manual verification is slow, not exhaustive and does not scale.\\

Our starting observation is that genuine certificates of the same issuer are produced by a stable generation pipeline and therefore share a technical fingerprint, in the same way that PDF producers can be recognized by their coding style \cite{5}. The key point is that this fingerprint is \emph{not} inferred from the file under investigation: it is learned beforehand from genuine reference certificates of known provenance, and its parameters are therefore ground-truth values. TrustPaper then answers a comparative question: \emph{is the submitted PDF consistent with the way its claimed issuer produces certificates?}

The contributions of this paper are:
\begin{itemize}
\item an explicit threat model for forged PDF certificates, with four attacker levels, that delimits what structural forensics can and cannot cover (Section~\ref{sec:threat});
\item an issuer-aware detection model that combines the claimed issuer, a reference fingerprint learned from genuine documents and an explainable aggregation of evidence, with a justified decision scale (Section~\ref{sec:method});
\item a controlled, reproducible evaluation of the open-source prototype that compares issuer-conditioned and issuer-agnostic analysis on the same forensic modules, per forgery technique and attacker level (Section~\ref{sec:eval}).
\end{itemize}

In this paper, the term issuer refers to the organization whose software generates the certificate, namely the teleconsultation platform. A standardized official form, by contrast, is considered a document type rather than an issuer, as it may be generated or completed using different tools. \\

The remainder of this paper is structured as follows. Section~\ref{sec:related} surveys related work in forged pdf detection in forensics. Section~\ref{sec:threat} defines the threat model and security objectives. Section~\ref{sec:method} presents our full methodology, including scoring model.  Section~\ref{sec:eval} details the experimental setup and evaluation metrics and discusses results and security implications. Finally, we conclude with future~directions.

\section{Related work }
\label{sec:related}
In literature, we could identify several vulnerabilities that have been exploited to circumvent existing PDF forgery detection mechanisms, thereby limiting their effectivenes:\\

\emph{Signature-centric analysis.} PDF security research showed that a valid signature does not imply that the content presented to a reader is trustworthy: signature wrapping and shadow attacks preserve a valid cryptographic status while changing the displayed content \cite{3}, \cite{4}. These works assume a signed document, whereas most certificates exchanged today are unsigned.\\

\emph{Generic structural scanners.} Tools such as PDFiD and ExifTool expose active content, metadata and revision traces \cite{6}, \cite{7}. They return raw indicators without a decision: whether an indicator is suspicious depends on how the legitimate document is normally produced, which these tools ignore.\\

\emph{Producer attribution.} Adhatarao and Lauradoux showed that the software that generated a PDF can be identified from its coding style \cite{5}. This answers \emph{which tool produced this file}, not \emph{is this file consistent with its claimed issuer}, and relies on a single family of features.\\

Our paper does not claim new low-level detectors; it reuses known PDF observations. Its contribution is to \emph{condition} several evidence families on the claimed issuer through a reference learned from genuine documents, and to combine them into an explainable score. Table~\ref{tab:positioning} summarizes the added value of our proposal, and Section~\ref{sec:eval} quantifies it.

\begin{table}[t]
\caption{Positioning with respect to existing approaches.}
\label{tab:positioning}
\centering
\scriptsize
\setlength{\tabcolsep}{3pt}
\begin{tabular}{L{0.38\columnwidth}cccc}
\toprule
Property & Sig.\ \cite{3,4} & Scan.\ \cite{6,7} & Style \cite{5} & Ours \\
\midrule
Applies to unsigned PDFs & \xmark & \cmark & \cmark & \cmark \\
Reference from genuine docs & \xmark & \xmark & \cmark & \cmark \\
Conditioned on claimed issuer & \xmark & \xmark & \xmark & \cmark \\
Combines evidence families & \xmark & \xmark & \xmark & \cmark \\
Observed vs.\ expected value & \xmark & \xmark & \xmark & \cmark \\
Decision (score and verdict) & \cmark & \xmark & \cmark & \cmark \\
\bottomrule
\end{tabular}
\end{table}

\begin{table}[t]
\caption{Attacker levels and expected coverage.}
\label{tab:attackers}
\centering
\scriptsize
\begin{tabular}{L{0.07\columnwidth}L{0.19\columnwidth}L{0.38\columnwidth}L{0.15\columnwidth}}
\toprule
Level & Knowledge & Typical manipulation & Coverage \\
\midrule
A0 & Generic editor & Edit and save, traces left & In scope \\
A1 & Metadata aware & Restore Producer, Creator, dates & In scope \\
A2 & Profile aware & Match fonts, revisions, QR syntax & Raise cost \\
A3 & Generator aware & Re-generate with issuer library & Out of scope \\
\bottomrule
\end{tabular}
\end{table}

\section{Threat model}
\label{sec:threat}
In this section, we formulate our threat model as an attack scenario, which we describe through the following key elements:\\

\emph{*) Assets and setting.} A relying party (employer, insurance fund) receives a PDF certificate that claims to come from an issuer $e$. The asset is the integrity of the certificate, i.e., that it was produced by $e$'s genuine pipeline and not modified afterwards. The defender has a set $R_e$ of genuine certificates of verified provenance, obtained from the issuer or from an independently recorded channel.\\

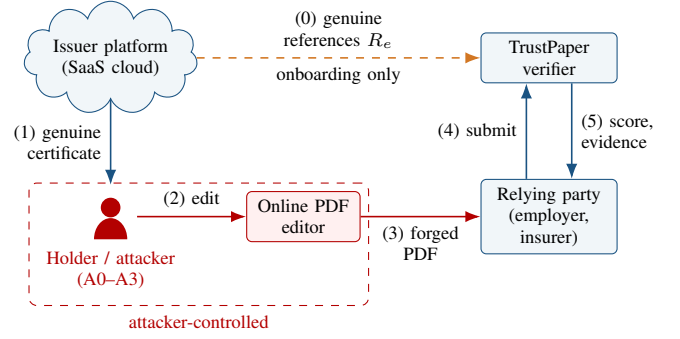
\begin{figure}[t]
\centering
\begin{tikzpicture}[font=\scriptsize]
  \node[cloud, draw=tpblue, fill=tplight, cloud puffs=11, aspect=2, inner sep=0pt, align=center] (issuer) at (0,2.3) {Issuer platform\\(SaaS cloud)};
  \node[tpbox, minimum width=18mm] (tp) at (5.8,2.3) {TrustPaper\\verifier};
  \pic at (0,-0.1) {person=red!70!black};
  \coordinate (attc) at (0,0.2);
  \node[align=center, text=red!70!black] (att) at (0,-0.5) {Holder / attacker\\(A0--A3)};
  \node[tpbox, draw=red!70!black, fill=red!6, minimum width=15mm] (ed) at (2.55,0.2) {Online PDF\\editor};
  \node[tpbox, minimum width=18mm] (rp) at (5.8,0.2) {Relying party\\(employer,\\insurer)};
  \draw[tparrow] (issuer.south) -- node[left, align=right]{(1) genuine\\certificate} (0,0.62);
  \draw[tparrow, draw=red!70!black] (0.35,0.2) -- node[above]{(2) edit} (ed.west);
  \draw[tparrow, draw=red!70!black] (ed.east) -- node[below, align=center]{(3) forged\\PDF} (rp.west);
  \draw[tparrow] ([xshift=-3mm]rp.north) -- node[left]{(4) submit} ([xshift=-3mm]tp.south);
  \draw[tparrow] ([xshift=3mm]tp.south) -- node[right, align=left]{(5) score,\\evidence} ([xshift=3mm]rp.north);
  \draw[tpdash] (issuer.east) -- node[above, align=center]{(0) genuine\\references $R_e$} node[below]{onboarding only} (tp.west);
  \node[draw=red!70!black, dashed, rounded corners=3pt, fit=(att)(ed)(attc), inner sep=3pt,
    label={[font=\scriptsize, text=red!70!black]below:attacker-controlled}] {};
\end{tikzpicture}
\caption{Attack scenario: The attacker edits a genuine certificate, or forges one from scratch, before submitting it to the relying party. The issuer only supplies genuine references (dashed) and does not take part in each verification.}
\label{fig:threat}
\end{figure}

\emph{*) Role of the issuer.} The issuer, typically a cloud teleconsultation platform, does not take part in the verification of each document. It intervenes only at onboarding and whenever its generator changes, by supplying genuine certificates or by validating the profile learned from them; it may also host the verification page targeted by its QR code. TrustPaper therefore decides without contacting the issuer, which keeps verification possible when the issuer offers no verification service. Fig.~\ref{fig:threat} summarizes the interactions.\\

\emph{*)Attacker.} The attacker starts either from a genuine certificate (to change dates, duration or identity) or from scratch, and controls every byte of the submitted file. We distinguish four levels of knowledge (Table~\ref{tab:attackers}). An opportunistic attacker (A0) uses an off-the-shelf editor without hiding its traces. A metadata-aware attacker (A1) additionally restores or rewrites metadata such as Producer, Creator and dates. A profile-aware attacker (A2) knows the fingerprint and reproduces its expected values, e.g., by reusing the fonts already present in the document. A generator-aware attacker (A3) recreates the PDF with the issuer's own library or serialization style. Here, we consider a grey-box attack scenario, where the adversary is assumed to have two distinct levels of knowledge about the target system.\\

Our TrustPaper is designed to detect A0 and A1 and to raise the cost of A2 and A3, and any attacker with access to the issuer's genuine pipeline, that can produce a structurally consistent false document.

\section{Issuer-aware detection model}
\label{sec:method}

\subsection{Overview}

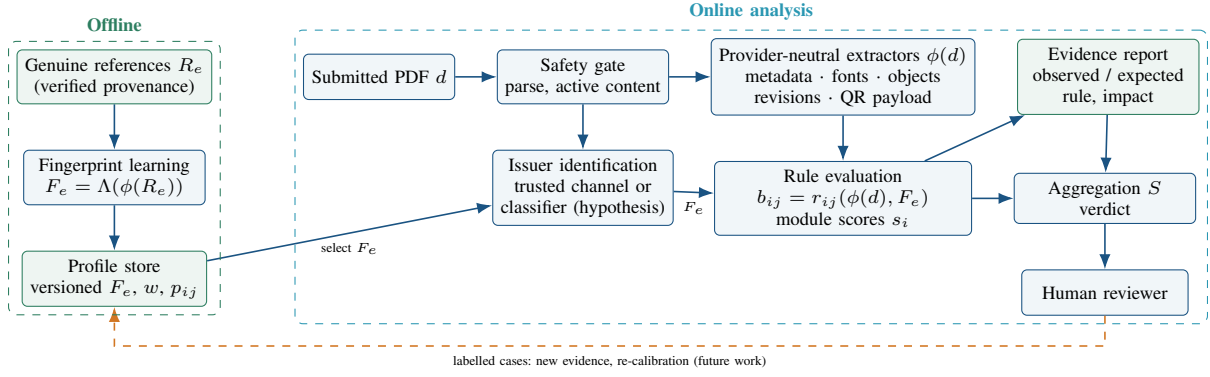
\begin{figure*}[t]
\centering
\begin{tikzpicture}[node distance=4mm and 5.5mm]
  \node[tpdata, minimum width=24mm] (refs) {Genuine references $R_e$\\(verified provenance)};
  \node[tpbox, minimum width=24mm, below=6mm of refs] (learn) {Fingerprint learning\\$F_e=\Lambda(\phi(R_e))$};
  \node[tpdata, minimum width=24mm, below=6mm of learn] (store) {Profile store\\versioned $F_e$, $w$, $p_{ij}$};
  \node[tpbox, minimum width=20mm, right=12mm of refs] (pdf) {Submitted PDF $d$};
  \node[tpbox, minimum width=20mm, right=of pdf] (gate) {Safety gate\\parse, active content};
  \node[tpbox, minimum width=34mm, right=of gate] (obs) {Provider-neutral extractors $\phi(d)$\\metadata $\cdot$ fonts $\cdot$ objects\\revisions $\cdot$ QR payload};
  \node[tpbox, minimum width=22mm, below=6mm of gate] (id) {Issuer identification\\trusted channel or\\classifier (hypothesis)};
  \node[tpbox, minimum width=34mm, below=6mm of obs] (rules) {Rule evaluation\\$b_{ij}=r_{ij}(\phi(d),F_e)$\\module scores $s_i$};
  \node[tpbox, minimum width=24mm, right=of rules] (agg) {Aggregation $S$\\verdict};
  \node[tpdata, minimum width=24mm, right=of obs] (rep) {Evidence report\\observed / expected\\rule, impact};
  \node[tpbox, minimum width=22mm, below=6mm of agg] (human) {Human reviewer};
  \draw[tparrow] (refs) -- (learn);
  \draw[tparrow] (learn) -- (store);
  \draw[tparrow] (pdf) -- (gate);
  \draw[tparrow] (gate) -- (obs);
  \draw[tparrow] (gate) -- (id);
  \draw[tparrow] (obs) -- (rules);
  \draw[tparrow] (store) -- node[below,font=\tiny]{select $F_e$} (id);
  \draw[tparrow] (id) -- node[below,font=\tiny]{$F_e$} (rules);
  \draw[tparrow] (rules) -- (agg);
  \draw[tparrow] (rules) -- (rep);
  \draw[tparrow] (agg) -- (human);
  \draw[tparrow] (rep) -- (agg);
  \draw[tpdash] (human.south) |- ++(0,-4mm) -| node[pos=0.25,below,font=\tiny]{labelled cases: new evidence, re-calibration (future work)} (store.south);
  \node[draw=tpgreen, dashed, rounded corners=3pt, fit=(refs)(learn)(store), inner sep=3pt,
    label={[font=\scriptsize\bfseries,text=tpgreen]above:Offline}] {};
  \node[draw=tpcyan, dashed, rounded corners=3pt, fit=(pdf)(gate)(obs)(id)(rules)(agg)(rep)(human), inner sep=3pt,
    label={[font=\scriptsize\bfseries,text=tpcyan]above:Online analysis}] {};
\end{tikzpicture}
\caption{Functional architecture: Observations acquire an issuer-specific meaning only through comparison with a fingerprint learned from genuine documents.}
\label{fig:archi}
\end{figure*}

Fig.~\ref{fig:archi} shows the functional architecture and how the modules interact. An \emph{offline} stage learns, for each issuer $e$, a reference fingerprint $F_e$ from genuine documents and stores it as a versioned declarative profile. The \emph{online} stage analyses a submitted PDF in four steps: (1) a safety gate rejects files that cannot be parsed or contain active content (JavaScript, launch actions); (2) provider-neutral extractors compute an observation vector; (3) the claimed issuer is determined and its fingerprint is selected; (4) rules compare observations with the fingerprint, and the resulting module scores are aggregated into a global score with an evidence report for a human reviewer.

\subsection{Reference fingerprint}

Let $\phi(d)=(\phi_1(d),\dots,\phi_K(d))$ be the provider-neutral observations of a document $d$: metadata strings, embedded font families and count, number of revisions, redefined objects, and QR payload. For an issuer $e$ with genuine references $R_e=\{d_1,\dots,d_n\}$, each dimension $k$ is summarized by an elementary rule that depends on its type:
\begin{align*}
\text{categorical: } & V_k=\textstyle\bigcup_{d\in R_e}\phi_k(d),\quad I_k=\bigcap_{d\in R_e}\phi_k(d),\\
\text{numeric: } & [m_k,M_k]=\big[\min_{d\in R_e}\phi_k(d),\ \max_{d\in R_e}\phi_k(d)\big],\\
\text{absence: } & a_k=\textstyle\bigwedge_{d\in R_e}\big(\phi_k(d)=\emptyset\big).
\end{align*}
The fingerprint is the collection $F_e=\{V_k,I_k,[m_k,M_k],a_k\}_k$, written $F_e=\Lambda(\phi(R_e))$. For fonts, for instance, $V_k$ contains every family seen in at least one reference (any other family is foreign), $I_k$ the families present in all references (a missing one is suspicious) and $[m_k,M_k]$ the range of the font count; for the modification date, $a_k$ is true when no reference has one. This is our formulation: an allow-list built only with union, intersection and range, so that every expected value can be traced back to genuine documents. Because $R_e$ contains only documents of verified provenance, these parameters are ground truth for the issuer. Table~\ref{tab:params} lists the parameters used, the rules derived from them and their impact.

\begin{table}[t]
\caption{Inputs of the score: module weight $w_i$, fingerprint parameters, derived rules and rule impact $p_{ij}$ (weights of the evaluated profiles).}
\label{tab:params}
\centering
\scriptsize
\setlength{\tabcolsep}{2.5pt}
\begin{tabular}{L{0.13\columnwidth}L{0.32\columnwidth}L{0.35\columnwidth}c}
\toprule
Module ($w_i$) & Learned on $R_e$ & Rule fires on $d$ if & $p_{ij}$ \\
\midrule
Import (0.10) & -- (gate) & Bad header or structure & reject \\
Malware (0.20) & -- (gate) & Active-content keyword & $20/$kw \\
\midrule
\multirow{4}{*}{\shortstack[l]{Metadata\\(0.35)}} & Producer values & Producer $\notin$ set & 70 \\
 & Creator values & Creator $\notin$ set & 55 \\
 & ModDate never present & ModDate present & 80 \\
 & No editor trace & Known editor in metadata & 85 \\
\midrule
\multirow{3}{*}{\shortstack[l]{Fonts\\(0.30)}} & Font families (union) & Foreign family & 65 \\
 & Families in every ref. & Required family missing & 65 \\
 & Font count range & Count above max / $\neq$ fixed & 55/50 \\
\midrule
Revis.\ (0.15) & Max.\ revision count & More revisions & 70 \\
Objects (0.10) & Max.\ redefined objects & More redefinitions & 45 \\
\midrule
\multirow{4}{*}{\shortstack[l]{QR code\\(0.10)}} & Present in every ref. & QR missing & 70 \\
 & Common URL prefix & URL outside issuer pattern & 60 \\
 & \multirow{2}{*}{Issue date} & Date in the future & 80 \\
 & & Issued $>$90 days ago & 45 \\
\bottomrule
\end{tabular}
\end{table}

Fingerprint and scoring parameters are stored in a versioned profile with a common schema for all issuers, so onboarding an issuer means collecting and validating $R_e$, not changing code. Listing~\ref{lst:yaml} shows an excerpt of the profile learned for issuer $P_A$ of Section~\ref{sec:eval}: the reference values come from $\Lambda$, while weights and impacts follow Table~\ref{tab:params}.

\begin{lstlisting}[language=yaml,frame=single,caption={Excerpt of the profile learned from 20 genuine references of $P_A$.},label={lst:yaml},float=t]
provider: p_a
version: learned-1.0
references:
  metadata: {expected_producers: [pypdf]}
  fonts:    {expected_fonts: [courier, helvetica],
             required_fonts: [courier, helvetica]}
steps:
  - module: metadata
    weight: 0.35
    rules:
      - {check: unexpected_producer, score_impact: 70}
      - {check: has_mod_date,        score_impact: 80}
  - module: fonts
    weight: 0.30
    rules:
      - {check: unexpected_font,     score_impact: 65}
      - {check: font_count_exceeded, score_impact: 55,
         threshold: 3}
\end{lstlisting}

\subsection{Scoring model}

For rule $j$ of module $i$, $b_{ij}\in\{0,1\}$ indicates whether the rule fired and $p_{ij}\in[0,100]$ is its impact. The module score is the strongest triggered rule,
\begin{equation}
s_i=\max_j\, b_{ij}\,p_{ij}.
\end{equation}
Rules of the same module usually observe the same edit (e.g., an injected font both adds a family and increases the count); taking the maximum avoids counting one manipulation several times. Independent evidence is combined \emph{across} modules by the global score
\begin{equation}
S = \max\left(\frac{\sum_i w_i s_i}{\sum_i w_i},\; \lambda \max_i s_i\right),\qquad \lambda=0.9.
\end{equation}
The first term rewards concordant evidence from several modules; the second prevents a single decisive anomaly from being diluted by modules that did not fire. This aggregation is our design choice. The verdict follows the bands 0--49, 50--69 and 70--100 of the prototype:
\begin{equation}
V(S)=\begin{cases}
\textit{authentic} & \text{if } S\le49,\\
\textit{suspicious} & \text{if } 49<S\le69,\\
\textit{falsification} & \text{if } S>69,
\end{cases}
\end{equation}
where \emph{authentic} means consistent with the issuer reference, not proven genuine.\\

Equation~(2) does not take raw observations: it takes, for every module $i$ executed on the document, its score $s_i\in[0,100]$ and its weight $w_i$, plus the constant $\lambda$ (Table~\ref{tab:params}). The modules are the two safety gates and the five forensic modules. The import gate passes ($s_i=0$) or rejects the file; the malware gate passes, rejects critical active content, or scores 20 per suspicious keyword (at most 95). The five forensic modules are metadata, fonts, revisions, objects and QR code; for them, $s_i$ is given by Eq.~(1) from the rules of the issuer profile. Font families, producer strings, revision counts or QR payloads thus enter the score only through the rules they trigger. With the evaluated profiles the weights sum to $\sum_i w_i=1.30$.

 The impacts, $\lambda$ and the two thresholds are not independent choices: they define three evidence tiers. When a single rule of impact $p$ fires, the weighted term is at most $0.35\,p$, so $S=0.9\,p$. A \emph{weak} rule ($p\le54$, e.g., one more redefined object, $p=45$) is only corroborating: alone it gives $S\le48.6$, so it never flags a document by itself. A \emph{strong} rule ($55\le p\le76$, e.g., foreign font, unexpected producer, missing QR) alone yields $49.5\le S\le68.4$, i.e., \emph{suspicious} and human review. A \emph{critical} rule ($p\ge77$, e.g., a modification date on a generator that never writes one, or an editor name in the metadata) alone yields $S\ge69.3$, i.e., \emph{falsification}. The value of $\lambda$ follows from these tiers. Keeping a single impact-50 rule authentic requires $\lambda\le49/50=0.98$; sending a single impact-55 rule to review requires $\lambda>49/55\approx0.891$; keeping impact-70 anomalies (unexpected producer, extra revision, missing QR) suspicious rather than falsification requires $\lambda\le69/70$; letting an impact-80 anomaly reach falsification requires $\lambda>69/80$. Any $\lambda\in(0.891,0.98]$ yields the same verdicts, and $\lambda=0.9$ is the smallest round value in this interval. With $\lambda=1$, a single unexpected producer would mean falsification; with $\lambda=0.8$, not even an impossible modification date would ($0.8\times80=64$).

The same anomaly does not carry the same importance for every issuer: for an issuer that appends a signing revision, a second revision is normal and the revision rule only fires above the learned count. Weights and impacts are fixed in the profile before any evaluation.

\subsection{Algorithm and output}

Algorithm~\ref{alg:tp} summarizes the analysis, implemented in an open-source prototype. Its output is not only $S$: for every module, the report stores $s_i$, the triggered rule, the observed value and the expected value from $F_e$, so that a reviewer can verify why a document was flagged.

\begin{algorithm}[t]
\caption{Issuer-aware analysis of a document $d$}
\label{alg:tp}
\small
\begin{algorithmic}[1]
\Require PDF $d$, profiles $\{(F_e, w^e, \{p^e_{ij}\})\}_e$
\Ensure score $S$, verdict, evidence set $E$
\If{\Call{Unparsable}{$d$} \textbf{or} \Call{ActiveContent}{$d$}}
  \State \Return \textsc{Reject} \Comment{safety gate, no score}
\EndIf
\State $o \gets \phi(d)$ \Comment{provider-neutral observations}
\State $e \gets$ \Call{ClaimedIssuer}{$d, o$}
\If{$e$ unknown \textbf{or} low confidence}
  \State $e \gets$ \textit{generic} \Comment{intrinsic checks only}
\EndIf
\State $E \gets \emptyset$
\For{each module $i$ in profile of $e$}
  \State $s_i \gets 0$
  \For{each rule $r_{ij}$ with impact $p_{ij}$}
    \If{$r_{ij}(o, F_e)$} \Comment{rule fires}
      \State $s_i \gets \max(s_i, p^e_{ij})$
      \State $E \gets E \cup \{(i, r_{ij}, o_{ij}, F_{e,ij})\}$
    \EndIf
  \EndFor
\EndFor
\State $S \gets \max\big(\sum_i w^e_i s_i / \sum_i w^e_i,\ \lambda \max_i s_i\big)$ \Comment{gates included}
\State \Return $S$, \Call{Verdict}{$S$}, $E$
\end{algorithmic}
\end{algorithm}

\emph{Issuer identification.} The claimed issuer should come from a trusted channel; otherwise a classifier proposes it from the text and metadata hints. Since these fields are attacker-controlled, this is a routing hypothesis, not an identity proof. Unknown issuers fall back to a generic profile that makes no issuer-consistency claim.

\section{Experimental validation}
\label{sec:eval}

\subsection{Protocol}

Real certificates contain health data and cannot be published. We therefore built a controlled, reproducible corpus with known ground truth that reproduces the generation patterns observed on real teleconsultation platforms. Two synthetic issuers are used. $P_A$ produces single-revision PDFs rewritten by a PDF library, with standard fonts, no modification date and a verification QR code. $P_B$ produces PDFs with an embedded TrueType font and an HTML-to-PDF producer string, then appends a signing revision, which is a legitimate incremental update.

For each issuer we generated several reference documents, used only to learn $F_e$, and disjoint genuine test documents. Each test document was then forged with the six techniques of Table~\ref{tab:results}, all targeting the issue date: F1 edits the date with an editor that embeds its own font and saves incrementally; F2 performs the same edit with a full re-export that writes the editor's producer and a modification date; F3 performs the edit with a full rewrite and restores the original metadata; F4 removes the verification QR code and F5 replaces it with a code pointing to another domain, both with restored metadata; F6 rewrites the date using the font resource already used by the page, with restored metadata. Impacts and weights were fixed before the experiment. 
(Table~\ref{tab:params}, weights 0.35, 0.30, 0.10, 0.15, 0.10 for metadata, fonts, objects, revisions and QR, plus 0.10 and 0.20 for the gates), and no parameter was tuned on the test documents. The claimed issuer is assumed to come from a trusted channel, so routing errors are not measured here.

Every document is scored twice by the same forensic modules: with the learned issuer fingerprint (\emph{issuer-aware}) and with the generic profile of the prototype (\emph{issuer-agnostic}), which applies fixed intrinsic rules such as ``any incremental update'' or ``any producer known as an editing or conversion tool''. This isolates the contribution of the reference layer; all scripts are released with the prototype.

\subsection{Results}

Table~\ref{tab:results} reports the documents flagged ($S>49$), the score per issuer and the modules that fired; rules being deterministic, all documents of a scenario obtained the same outcome.

\begin{table}[t]
\caption{Detection per forgery technique (60 documents per row, 30 per issuer). Flagged: $S>49$. Modules: M metadata, F fonts, O objects, R revisions, Q QR code.}
\label{tab:results}
\centering
\scriptsize
\setlength{\tabcolsep}{2.2pt}
\begin{tabular}{llcccccc}
\toprule
 & & \multicolumn{4}{c}{Issuer-aware} & \multicolumn{2}{c}{Issuer-agnostic} \\
\cmidrule(lr){3-6}\cmidrule(lr){7-8}
ID & Lvl & Flagged & $S$ ($P_A$) & $S$ ($P_B$) & Modules & Flagged & $S$ \\
\midrule
G  & --  & 0/60  & 0.0  & 0.0  & --    & 60/60 & 49.5 \\
F1 & A0 & 60/60 & 63.0 & 63.0 & F,R (+O for $P_A$) & 60/60 & 49.5 \\
F2 & A0 & 60/60 & 76.5 & 63.0 & M,F   & 60/60 & 49.5 \\
F3 & A1 & 60/60 & 58.5 & 58.5 & F     & 60/60 & 49.5 \\
F4 & A1 & 60/60 & 63.0 & 63.0 & Q     & 60/60 & 49.5 \\
F5 & A1 & 60/60 & 54.0 & 54.0 & Q     & 60/60 & 49.5 \\
F6 & A2 & 0/60  & 0.0  & 0.0  & --    & 60/60 & 49.5 \\
\bottomrule
\end{tabular}
\end{table}

\emph{Issuer-aware analysis.} All 300 forgeries of levels A0 and A1 are flagged and none of the 60 genuine test documents is, which obtained a score of 0. Each technique is caught by the module designed for it (fonts for F1--F3, revisions for F1, metadata for F2, QR for F4--F5). The issuers also show issuer-specific interpretation: F1 triggers the object rule for $P_A$, which never redefines objects, but not for $P_B$, whose signing revision legitimately redefines two; and the editor-trace rule is absent from $P_B$'s fingerprint because its legitimate producer is itself in the generic tool list, so F2 scores 63 instead of 76.5.

\emph{Case studies.} Table~\ref{tab:cases} shows the evidence report of four test documents as returned by the engine. For F2, Eq.~(2) gives $S=\max\big((0.35\times85+0.30\times65)/1.30,\;0.9\times85\big)=\max(37.9,76.5)=76.5$, hence \emph{falsification}. In F3, restored metadata hides the editor but the foreign font still exposes the edit.

\begin{table}[t]
\caption{Case studies: evidence reported by the engine (observed vs.\ expected value), module scores and verdict.}
\label{tab:cases}
\centering
\scriptsize
\setlength{\tabcolsep}{2.5pt}
\begin{tabular}{L{0.11\columnwidth}L{0.47\columnwidth}L{0.17\columnwidth}L{0.16\columnwidth}}
\toprule
Case & Evidence (observed / expected) & $s_i$ & $S$, verdict \\
\midrule
F2, $P_A$ & Producer Canva / pypdf; ModDate present / never; family DejaVu Sans / \{Courier, Helvetica\}; 4 fonts / 3 & M 85, F 65 & 76.5, falsif. \\
F3, $P_A$ & Metadata restored (pypdf, no ModDate); family DejaVu Sans / \{Courier, Helvetica\}; 4 fonts / 3 & F 65 & 58.5, susp. \\
F4, $P_A$ & QR code absent / present in every reference & Q 70 & 63.0, susp. \\
F1, $P_B$ & Family DejaVu Sans / \{Vera, Helvetica\}; 3 revisions / max.\ 2; 2 redefined objects / max.\ 2 (no rule) & F 65, R 70 & 63.0, susp. \\
\bottomrule
\end{tabular}
\end{table}

\emph{Issuer-agnostic baseline.} Without a reference, the same modules give every document, genuine or forged, a score of 49.5, hence the verdict \emph{suspicious}. Generic rules fire on legitimate constructions (unreferenced objects left by $P_A$'s library; $P_B$'s signing revision and HTML-to-PDF producer) as well as on forgeries, at the same capped impact. The baseline therefore cannot separate the two classes: whatever the band limit, it flags either all genuine documents and all forgeries, or none of them (Table~\ref{tab:sweep}).

\begin{table}[t]
\caption{Sensitivity to the upper limit $b$ of the authentic band (flag if $S>b$; the prototype uses $b=49$): detection rate on A0--A1 forgeries (TPR) and false-positive rate on genuine documents (FPR), in \%.}
\label{tab:sweep}
\centering
\scriptsize
\begin{tabular}{ccccc}
\toprule
 & \multicolumn{2}{c}{Issuer-aware} & \multicolumn{2}{c}{Issuer-agnostic} \\
\cmidrule(lr){2-3}\cmidrule(lr){4-5}
$b$ & TPR & FPR & TPR & FPR \\
\midrule
39 & 100 & 0 & 100 & 100 \\
49 & 100 & 0 & 100 & 100 \\
59 & 60 & 0 & 0 & 0 \\
69 & 10 & 0 & 0 & 0 \\
\bottomrule
\end{tabular}
\end{table}

\emph{Adaptive attacker.} F6 (A2) is rated authentic by the issuer-aware analysis: the date is rewritten with a font resource that already belongs to the page, the file is fully rewritten and the original metadata is restored, so every learned parameter is matched. This is consistent with the threat model, which places A2 at the boundary of structural forensics.

\emph{Decision scale and cost.} Most A0--A1 forgeries are rated \emph{suspicious}; only critical anomalies reach \emph{falsification}. With the current weights, the weighted term never exceeded the dominance term, so concordant evidence (F1, F2) did not raise the score beyond its strongest anomaly; additive or probabilistic combinations calibrated on labelled data are a natural improvement. The median analysis time (safety gates and forensic modules) was 658\,ms.

\section{Discussion and limitations}
\label{sec:discussion}

TrustPaper detects inconsistency with a reference, not semantic truth. A2 and A3 attackers, and a false statement produced through the genuine pipeline, remain undetected. Harder-to-mimic dimensions (content-stream operator distributions, coordinate precision, resource ordering, cross-reference layout) would raise the cost of A2, and issuer-signed payloads should take precedence when available.

A profile is a security asset: a legitimate generator upgrade raises alerts until a new profile version is approved, and a tampered profile would bias every decision. Low-confidence issuer routing should abstain rather than silently apply an unrelated profile.

The synthetic corpus shows that each forgery technique triggers the module designed for it and that the reference changes the decision, but it does not estimate real-world rates. An evaluation on real certificates, with source-disjoint splits, blinded labelling and a documented legal basis, is the next step.

A standardized official form is a template filled by many tools, not an issuer. Its fingerprint only covers template invariants (structure, fields, layout); e.g., a template authored with desktop-publishing software would wrongly trigger generic producer rules. It thus receives a reduced document-type profile.

\emph{Toward evidence learning.} Documents confirmed as genuine or forged by reviewers can be used to calibrate impacts and thresholds on real data and to discover new evidence for each issuer (dashed loop in Fig.~\ref{fig:archi}), while preserving per-rule explanations. Another direction for future
improvement would be to incorporate AI-based approaches to
improve detection accuracy and overall system effectiveness
through data-driven learning.

\section{Conclusion}

TrustPaper assesses whether a PDF certificate is consistent with the way its claimed issuer produces documents, by combining the claimed issuer, a fingerprint learned from genuine documents and an explainable aggregation of evidence. A controlled evaluation shows that this reference layer turns ambiguous PDF observations into a usable decision against opportunistic and metadata-aware forgers. Several future work are presented such as evaluating the proposed approach on real certificates from different platforms, add harder-to-mimic features and calibrate the scoring using data-driven learning.


\begin{thebibliography}{1}

\bibitem{1}
Assurance Maladie, ``Lutte contre la fraude à l'Assurance Maladie : bilan annuel'', https://www.ameli.fr (accessed Sep. 2026).
\bibitem{2}
ISO 32000-2:2020, ``Document management -- Portable document format -- Part 2: PDF 2.0'', International Organization for Standardization, 2020.
\bibitem{3}
V. Mladenov, C. Mainka, K. Meyer zu Selhausen, M. Grothe and J. Schwenk, ``1 Trillion Dollar Refund: How To Spoof PDF Signatures'', in \emph{Proc. ACM SIGSAC Conf. Computer and Communications Security (CCS)}, London, UK, 2019.
\bibitem{4}
C. Mainka, V. Mladenov and S. Rohlmann, ``Shadow Attacks: Hiding and Replacing Content in Signed PDFs'', in \emph{Proc. Network and Distributed System Security Symp. (NDSS)}, 2021.
\bibitem{5}
S. Adhatarao and C. Lauradoux, ``Robust PDF Files Forensics Using Coding Style'', in \emph{Proc. IFIP Int. Conf. ICT Systems Security and Privacy Protection (SEC)}, 2022.
\bibitem{6}
D. Stevens, ``PDFiD'', \texttt{blog.didierstevens.com/programs/pdf-tools} (accessed Sep. 2026).
\bibitem{7}
P. Harvey, ``ExifTool'', https://exiftool.org (accessed Sep. 2026).
\bibitem{8}
Boonkrong, Sirapat. "Design of an academic document forgery detection system." International Journal of Information Technology 17.9 (2025): 5175-5187.
\bibitem{9}
Sukhija, Ranveer, Munish Kumar, and Manish Kumar Jindal. "Document forgery detection: a comprehensive review." International Journal of Data Science and Analytics 20.5 (2025): 4385-4407.
\end{thebibliography}
\end{document}